\documentclass[12pt]{article}

\usepackage[totalwidth=460truept,totalheight=620truept]{geometry}
\usepackage{amsfonts,latexsym,amssymb,amsmath,graphicx,accents,slashed,subfigure}
\usepackage[T1]{fontenc}
\usepackage[hidelinks]{hyperref}

\global\arraycolsep=1truept

\numberwithin{equation}{section}

\begin{document}

\bigskip \phantom{C}

\vskip 1.5truecm

\begin{center}
{\huge \textbf{The Ultraviolet Behavior}}

\vskip.4truecm

{\huge \textbf{Of Quantum Gravity}}

\vskip1truecm

\textsl{Damiano Anselmi\footnote{%
damiano.anselmi@unipi.it} and Marco Piva\footnote{%
marco.piva@df.unipi.it}}

\vskip .1truecm

\textit{Dipartimento di Fisica ``Enrico Fermi'', Universit\`{a} di Pisa, }

\textit{Largo B. Pontecorvo 3, 56127 Pisa, Italy}

\textit{and INFN, Sezione di Pisa,}

\textit{Largo B. Pontecorvo 3, 56127 Pisa, Italy}

\vskip1truecm

\textbf{Abstract}
\end{center}

A theory of quantum gravity has been recently proposed by means of a novel
quantization prescription, which is able to turn the poles of the free
propagators that are due to the higher derivatives into fakeons. The
classical Lagrangian contains the cosmological term, the Hilbert term, $%
\sqrt{-g}R_{\mu \nu }R^{\mu \nu }$ and $\sqrt{-g}R^{2}$. In this paper, we
compute the one-loop renormalization of the theory and the absorptive part
of the graviton self energy. The results illustrate the mechanism that makes
renormalizability compatible with unitarity. The fakeons disentangle the
real part of the self energy from the imaginary part. The former obeys a
renormalizable power counting, while the latter obeys the nonrenormalizable
power counting of the low energy expansion and is consistent with unitarity
in the limit of vanishing cosmological constant. The value of the absorptive
part is related to the central charge $c$ of the matter fields coupled to
gravity.

\vfill\eject

\section{Introduction}

\label{intro} \setcounter{equation}{0}

The problem of quantum gravity is the difficulty to reconcile
renormalizability and perturbative unitarity. A solution has been recently
proposed \cite{LWgrav} by turning the ghosts due to the higher derivatives
into \textit{fakeons} \cite{fakeons}, or fake degrees of freedom, which
contribute to the correlation functions, but disappear from the physical
spectrum. The idea amounts to a novel prescription to treat the poles of the
free propagators. It is suggested by the reformulation of the Lee-Wick
models \cite{leewick} as nonanalytically Wick rotated Euclidean theories 
\cite{LWformulation,LWunitarity}. The fakeons clarify the properties of
these models and refine their original, incomplete formulation. At the same
time, they have a broader range of applications, to the extent that they can
be introduced in models that are not of the Lee-Wick type.

Several options for quantum gravity emerge from this approach. A unique one
among them is strictly renormalizable \cite{LWgrav}. Its classical
Lagrangian contains the cosmological term $\sqrt{-g}$, the Hilbert term $%
\sqrt{-g}R$ and the quadratic terms $\sqrt{-g}R_{\mu \nu }R^{\mu \nu }$ and $%
\sqrt{-g}R^{2}$. The other options are infinitely many and
super-renormalizable, which makes them less attractive from the physical
point of view. The proof of perturbative unitarity can be carried out to the
very end \cite{fakeons}, once the effects of the cosmological constant are
neglected. The reason is that a satisfactory scattering theory has not been
developed, yet, in the presence of a cosmological constant (see refs. \cite%
{adsscatt} for some investigations and proposals on this issue) and it might
even not exist. If that were the case, a nonvanishing cosmological constant
(generically turned on by the radiative corrections) would signal a
unitarity anomaly in the universe, which would explain why this quantity is
so small. Super-renormalizable theories of quantum gravity where the
cosmological constant is not turned on by the radiative corrections can be
built \cite{LWgrav}, yet it is hard to argue that they describe the laws of
physics.

We think that the strictly renormalizable option is at present the best
candidate to explain quantum gravity, even if the cosmological constant
cannot be turned off to all orders. In this paper, we compute the key
quantities of this theory at one loop. Specifically, we work out the
absorptive part of the graviton self energy and the one-loop
renormalization, both in the pure theory and in the presence of matter.

It can be shown \cite{LWgrav,fakeons} that the fakeon prescription does not
affect the renormalization, which coincides with the one of the Euclidean
version of the theory. The one-loop beta functions can be calculated by
working out the divergent parts of the two graviton and three graviton
correlation functions. However, some diagrams with three external gravitons
are very involved. Their computation can be avoided by means of the
Ward-Takahashi-Slavnov-Taylor (WTST)\ identities \cite{WTST}. The most
popular technique to achieve this goal is the background field method \cite%
{backfm,percacci}, which incorporates the WTST identities by gauge fixing
the theory in a clever way. A more standard approach is the one pursued by
Salvio and Strumia in ref. \cite{agravity,agravity2}. They replace the
computations of the diagrams with three external gravitons with the
computations of the (much simpler) diagrams with one external graviton and
two external matter fields.

In the first part of this paper, we compute the one-loop renormalization of
the theory by means of a third procedure, which does not make use of the
background field method and remains within the gravity sector. We quantize
the theory with the help of the Batalin-Vilkovisky formalism \cite{bata} and
calculate the divergences of the graviton self energy and those of the
diagrams that renormalize the symmetry transformations of the fields. This
approach gives a few results that are not available in the literature, such
as the nonlinear contributions to the field redefinitions of the metric
tensor and the Faddeev-Popov ghosts.

Then we turn to the calculation of the absorptive part of the graviton self
energy, which involves novel techniques, suggested by the properties of the
fakeons. For simplicity, we work in the limit of vanishing cosmological
constant. A number of tricks allow us to relate the absorptive part to the
renormalization of the low-energy theory, obtained by expanding the action
around the Hilbert term and treating the higher-derivative terms
perturbatively. The calculation obeys the power counting of the ordinary,
nonrenormalizable Einstein theory, but its outcome is convergent and
consistent with unitarity, by a peculiar mechanism due to the fakeons. The
final formula of the absorptive part is piecewise local, equal to a
contribution due to the so-called \textit{central charge} $c$ of the matter
fields coupled to gravity plus a correction due to the nonminimal couplings
of the scalar fields plus terms that vanish on the solutions of the field
equations. The results show that the quantum gravity theory of ref. \cite%
{LWgrav} gives physical predictions that differ from those of any other
quantization of the same classical action \cite%
{backfm,percacci,agravity,agravity2,stelle}.

We use the dimensional regularization. The paper is organized as follows. In
section \ref{QG} we quantize the theory to the extent that is strictly
necessary for the calculation of the one-loop renormalization. In section %
\ref{reno} we work out the beta functions at one loop and the
renormalizations of the fields. In section \ref{abso} we complete the
quantization of the theory by detailing the graviton/fakeon prescription for
the propagators. Then we calculate the absorptive part of the graviton self
energy by relating it to the renormalization of the theory expanded around
the Hilbert term. Section \ref{concl} contains the conclusions and an
outlook about generalizations of the calculations performed here.

\section{Quantum gravity in the Batalin-Vilkovisky formalism}

\setcounter{equation}{0} \label{QG}

The strictly renormalizable theory of quantum gravity proposed in ref. \cite%
{LWgrav} has action%
\begin{equation}
S_{\text{HD}}=-\frac{\mu ^{-\varepsilon }}{2\kappa ^{2}}\int \sqrt{-g}\left[
2\Lambda _{C}+\zeta R+\alpha \left( R_{\mu \nu }R^{\mu \nu }-\frac{1}{3}%
R^{2}\right) -\frac{\xi }{6}R^{2}\right] ,  \label{lhd}
\end{equation}%
where $\alpha $, $\xi $, $\zeta $, $\Lambda _{C}$ and $\kappa $ are real
constants, with $\alpha >0$, $\xi >0$ and $\zeta >0$, while $\mu $ is the
dynamical scale and $\varepsilon =4-D$, $D$ being the continued spacetime
dimension introduced by the dimensional regularization. The action (\ref{lhd}%
) is quantized by means of a novel \textit{graviton/fakeon prescription},
which is formulated in detail in section \ref{abso}. We skip this part for
the time being, because we want to concentrate on the one-loop
renormalization, which coincides with the one of the Euclidean version of
the theory \cite{LWgrav,fakeons}.

To apply the procedure described in the introduction and handle the WTST
identities in a compact form, we use the Batalin-Vilkovisky formalism \cite%
{bata}, which is a formal refinement of the Zinn-Justin approach \cite%
{zinnjustin}. We collect the fields into the row%
\begin{equation*}
\Phi ^{\alpha }=\{g_{\mu \nu },C^{\rho },\bar{C}^{\sigma },B^{\tau }\},
\end{equation*}%
where $C^{\rho }$ and $\bar{C}^{\sigma }$ are the Faddeev-Popov ghosts and
antighosts of diffeomorphisms, respectively, while $B^{\tau }$ are the
Lagrange multipliers for the gauge fixing (also known as Nakanishi-Lautrup
fields \cite{nakalau}). We also introduce a row of external sources 
\begin{equation*}
K_{\alpha }=\{K^{\mu \nu },K_{\rho }^{C},K_{\sigma }^{\bar{C}},K_{\tau
}^{B}\},
\end{equation*}%
conjugate to the fields, and define the \textit{antiparentheses} of two
functionals $X$ and $Y$ of $\Phi $ and $K$ as 
\begin{equation*}
(X,Y)\equiv \int \left( \frac{\delta _{r}X}{\delta \Phi ^{\alpha }}\frac{%
\delta _{l}Y}{\delta K_{\alpha }}-\frac{\delta _{r}X}{\delta K_{\alpha }}%
\frac{\delta _{l}Y}{\delta \Phi ^{\alpha }}\right) ,
\end{equation*}%
where the integral is over the spacetime points associated with repeated
indices and the subscripts $l$, $r$ in $\delta _{l}$, $\delta _{r}$ denote
the left and right functional derivatives, respectively.

The next step is to extend the action $S_{\text{HD}}$ into 
\begin{equation}
S(\Phi ,K)=S_{\text{HD}}+(S_{K},\Psi )+S_{K},  \label{sfik}
\end{equation}%
where $\Psi (\Phi )$ is a functional of the fields, called \textit{gauge
fermion}, which is used to fix the gauge, while 
\begin{equation*}
S_{K}=-\int \mathcal{R}^{\alpha }(\Phi )K_{\alpha }=\int (g_{\mu \rho
}\partial _{\nu }C^{\rho }+g_{\nu \rho }\partial _{\mu }C^{\rho }+C^{\rho
}\partial _{\rho }g_{\mu \nu })K^{\mu \nu }+\int C^{\sigma }(\partial
_{\sigma }C^{\rho })K_{\rho }^{C}-\int B^{\sigma }K_{\sigma }^{\bar{C}}
\end{equation*}%
collects the infinitesimal symmetry transformations $\mathcal{R}^{\alpha
}(\Phi )$ of the fields, coupled to the sources $K_{\alpha }$. In
particular, the functions%
\begin{equation*}
-\frac{\delta _{r}S}{\delta K^{\mu \nu }}=\mathcal{R}_{\mu \nu }(g,C)\equiv
-g_{\mu \rho }\partial _{\nu }C^{\rho }-g_{\nu \rho }\partial _{\mu }C^{\rho
}-C^{\rho }\partial _{\rho }g_{\mu \nu }
\end{equation*}%
are inherited from the infinitesimal transformations $\delta _{\Sigma
}g_{\mu \nu }=\mathcal{R}_{\mu \nu }(g,\Sigma )$ of the metric tensor $%
g_{\mu \nu }$ under diffeomorphisms, where $\Sigma ^{\rho }$ are functions
of the spacetime point.

The action (\ref{sfik}) satisfies the \textit{master equation} (also known
as Zinn-Justin equation)%
\begin{equation}
(S,S)=0,  \label{mast}
\end{equation}%
which collects the gauge invariance of $S_{\text{HD}}$ and the closure of
the symmetry transformations. The generating functional $Z$ of the
correlation functions and the generating functional $W$ of the connected
correlation functions are defined by the formulas 
\begin{equation*}
Z(J,K)=\int [\mathrm{d}\Phi ]\exp \left( iS(\Phi ,K)+i\int \Phi ^{\alpha
}J_{\alpha }\right) =\exp iW(J,K).
\end{equation*}%
The \textquotedblleft quantum effective action\textquotedblright , i.e. the
generating functional $\Gamma (\Phi ,K)=W(J,K)-\int \Phi ^{\alpha }J_{\alpha
}$ of the one-particle irreducible diagrams, is defined as the Legendre
transform of $W(J,K)$ with respect to $J$, where $\Phi ^{\alpha }=\delta
_{r}W/\delta J_{\alpha }$. It is easy to see that (\ref{mast}) implies that $%
\Gamma $ satisfies an analogous master equation%
\begin{equation}
(\Gamma ,\Gamma )=0,  \label{mastg}
\end{equation}%
which collects all the WTST identities in a compact form.

By renormalizing the action (\ref{sfik}) and taking advantage of the
properties of the Batalin-Vilkovisky formalism, we can work out the beta
functions without computing the renormalization of the three-graviton vertex
and without introducing matter fields. It is sufficient to renormalize $%
S_{K} $ (which is relatively easy) and the graviton self energy (which is
more demanding).

We expand the metric tensor $g_{\mu \nu }$ around the flat-space metric $%
\eta _{\mu \nu }=$ diag$(1,-1,-1,-1)$ by writing%
\begin{equation*}
g_{\mu \nu }=\eta _{\mu \nu }+2\kappa h_{\mu \nu },
\end{equation*}%
where $h_{\mu \nu }$ is the quantum fluctuation. We further define $h\equiv
\eta ^{\mu \nu }h_{\mu \nu }$. The indices of $\partial _{\mu }$, $h_{\mu
\nu }$, the fields $\Phi ^{\alpha }$\ (except $g_{\mu \nu }$) and the
sources $K_{\alpha }$ are raised and lowered by means of the flat-space
metric. We raise and lower the indices of the covariant derivatives, the
metric $g_{\mu \nu }$, the Riemann tensor and the Ricci tensor by means of $%
g_{\mu \nu }$.

We choose the gauge fermion%
\begin{equation*}
\Psi =\mu ^{-\varepsilon }\int \bar{C}^{\mu }\left( \sigma \zeta +\alpha
\square \right) \left( \mathcal{G}_{\mu }-\frac{\kappa ^{2}}{\lambda }B_{\mu
}\right) ,
\end{equation*}%
where $\square =\eta ^{\mu \nu }\partial _{\mu }\partial _{\nu }$ is the
flat-space D'Alembertian,%
\begin{equation}
\mathcal{G}_{\mu }(g)=\eta ^{\nu \rho }\partial _{\rho }g_{\mu \nu }-(\omega
+1)\eta ^{\nu \rho }\partial _{\mu }g_{\nu \rho }=2\kappa \lbrack \partial
_{\nu }h_{\mu }^{\nu }-(\omega +1)\partial _{\mu }h]  \label{gmu}
\end{equation}%
is the gauge-fixing function and $\sigma $, $\lambda $ and $\omega $ are
gauge-fixing parameters.

The gauge-fixed action reads%
\begin{equation}
S_{\text{gf}}=S_{\text{HD}}+(S_{K},\Psi ),  \label{lgr}
\end{equation}%
where%
\begin{equation}
(S_{K},\Psi )=\mu ^{-\varepsilon }\int B^{\mu }\left( \sigma \zeta +\alpha
\square \right) \left( \mathcal{G}_{\mu }-\frac{\kappa ^{2}}{\lambda }B_{\mu
}\right) +S_{\text{gh}}  \label{skpsi}
\end{equation}%
and the action $S_{\text{gh}}$ of the Faddeev-Popov ghosts reads%
\begin{equation}
S_{\text{gh}}=\mu ^{-\varepsilon }\int \left[ \bar{C}^{\mu }\partial ^{\nu
}-(\omega +1)\eta ^{\mu \nu }\bar{C}^{\tau }\partial _{\tau }\right] \left(
\sigma \zeta +\alpha \square \right) \left[ g_{\mu \rho }\partial _{\nu
}C^{\rho }+g_{\nu \rho }\partial _{\mu }C^{\rho }+C^{\rho }\partial _{\rho
}g_{\mu \nu }\right] .  \label{ghac}
\end{equation}%
If we make the field redefinition $\bar{C}^{\prime \mu }=\left( \sigma \zeta
+\alpha \square \right) \bar{C}^{\mu }$ on the antighosts, the ghost action
turns into the more conventional form%
\begin{equation}
S_{\text{gh}}=\mu ^{-\varepsilon }\int \left[ \bar{C}^{\prime \mu }\partial
^{\nu }-(\omega +1)\eta ^{\mu \nu }\bar{C}^{\prime \tau }\partial _{\tau }%
\right] \left[ g_{\mu \rho }\partial _{\nu }C^{\rho }+g_{\nu \rho }\partial
_{\mu }C^{\rho }+C^{\rho }\partial _{\rho }g_{\mu \nu }\right] .
\label{ghaccon}
\end{equation}%
The ghost actions (\ref{ghac}) and (\ref{ghaccon}) are equivalent for our
purposes of this paper.

\section{Renormalization}

\setcounter{equation}{0} \label{reno}

In this section we calculate the renormalization of the theory at one loop.
Let $S_{\text{count}}$ denote the one-loop counterterm action. A few
standard properties allow us to give $S_{\text{count}}$ an explicit form.
First, it is easy to show that the master equations (\ref{mast}) and (\ref%
{mastg}) imply the identity%
\begin{equation}
(S,S_{\text{count}})=0.  \label{coho}
\end{equation}%
Second, $S_{\text{count}}$ cannot depend on $B$, $K^{\bar{C}}$ and $K^{B}$,
because no vertices of the action (\ref{sfik}) contain them, so no
one-particle irreducible diagrams can be built with $B$, $K^{\bar{C}}$
and/or $K^{B}$ on the external legs. Third, $S$ does not depend on $K^{\mu
\nu }$ and $\bar{C}^{\rho }$ separately, but only through the combination 
\begin{equation*}
\tilde{K}^{\mu \nu }=K^{\mu \nu }+\mu ^{-\varepsilon }\left( \sigma \zeta
+\alpha \square \right) \int \frac{\delta \mathcal{G}_{\rho }}{\delta g_{\mu
\nu }}\bar{C}^{\rho },
\end{equation*}%
so the same is true of $S_{\text{count}}$.

On general grounds\footnote{%
A convenient way to prove formulas (\ref{scount}) and (\ref{ffk}) is by
interpolating back and forth between the background field approach and the
ordinary approach \cite{nocohoKSZ}.}, the solution of (\ref{coho}) can be
written as%
\begin{equation}
S_{\text{count}}=\frac{\mu ^{-\varepsilon }}{(4\pi )^{2}\varepsilon }\int 
\sqrt{-g}\left[ 2\Delta \Lambda _{C}+\Delta \zeta R+\Delta \alpha \left(
R_{\mu \nu }R^{\mu \nu }-\frac{1}{3}R^{2}\right) -\frac{\Delta \xi }{6}R^{2}%
\right] +(S,\mathcal{F}),  \label{scount}
\end{equation}%
where $\Delta \Lambda _{C}$, $\Delta \zeta $, $\Delta \alpha $ and $\Delta
\xi $ are constants and $\mathcal{F}(\Phi ,K)$ is a local functional of
ghost number minus one, equal to the integral of a local function of
dimension three. Using (\ref{coho}), it is easy to show that $\mathcal{F}$
also depends on $K^{\mu \nu }$ and $\bar{C}^{\rho }$ via the combination $%
\tilde{K}^{\mu \nu }$. Then, the dimension of $\mathcal{F}$ and its ghost
number imply that we can parametrize it as 
\begin{equation}
\mathcal{F}(\Phi ,K)=\int \Delta g_{\mu \nu }\tilde{K}^{\mu \nu }+\int
\Delta C^{\rho }K_{\rho }^{C},  \label{ffk}
\end{equation}%
where $\Delta g_{\mu \nu }$ and $\Delta C^{\rho }$ are the renormalizations
of the metric tensor and the Faddeev-Popov ghosts, respectively. They
generalize the more common multiplications by wave function renormalization
constants.

\begin{figure}[t]
\begin{center}
\includegraphics[width=16truecm]{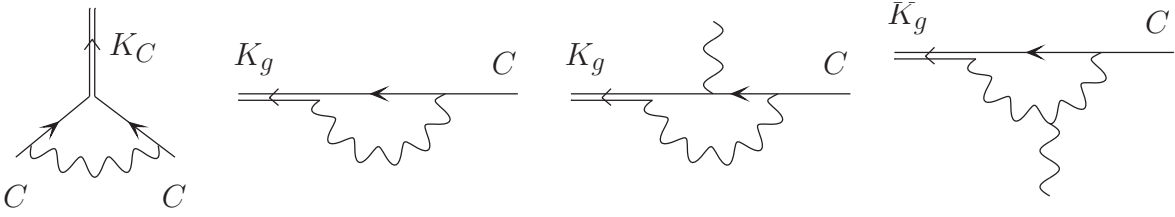}
\end{center}
\caption{Diagrams that renormalize the symmetry transformations of the
fields }
\label{grafici}
\end{figure}

A straightforward calculation gives%
\begin{equation}
(S,\mathcal{F})=\int \frac{\delta S_{\text{HD}}}{\delta g_{\mu \nu }}\Delta
g_{\mu \nu }-\int \Delta \mathcal{R}_{\mu \nu }\tilde{K}^{\mu \nu }-\int
\Delta \mathcal{R}^{\rho }K_{\rho }^{C},  \label{sf}
\end{equation}%
where%
\begin{eqnarray*}
\Delta \mathcal{R}_{\mu \nu } &=&-(S,\Delta g_{\mu \nu })+\int \Delta
g_{\alpha \beta }\frac{\delta _{l}\mathcal{R}_{\mu \nu }(g,C)}{\delta
g_{\alpha \beta }}+\int \Delta C^{\tau }\frac{\delta _{l}\mathcal{R}_{\mu
\nu }(g,C)}{\delta C^{\tau }}, \\
\Delta \mathcal{R}^{\rho } &=&-(S,\Delta C^{\rho })+\int \Delta C^{\tau }%
\frac{\delta _{l}\mathcal{R}^{\rho }(C)}{\delta C^{\tau }},
\end{eqnarray*}%
where $\mathcal{R}^{\rho }(C)=-\delta _{r}S/\delta K_{\rho }^{C}$. To the
quadratic order in the fields, we can parametrize the field redefinitions as 
\begin{eqnarray}
\Delta C^{\rho } &=&\kappa ^{2}s_{1}C^{\rho }+\kappa ^{3}s_{2}h_{\mu }^{\rho
}C^{\mu }+\kappa ^{3}s_{3}hC^{\rho },  \notag \\
\Delta g_{\mu \nu } &=&\kappa ^{2}t_{0}g_{\mu \nu }+\kappa ^{3}(t_{1}h_{\mu
\nu }+t_{2}\eta _{\mu \nu }h)  \notag \\
&&+\kappa ^{4}[t_{3}h_{\mu }^{\rho }h_{\rho \nu }+t_{4}hh_{\mu \nu }+\eta
_{\mu \nu }(t_{5}h_{\rho \sigma }h^{\rho \sigma }+t_{6}h^{2})],
\label{fieldredef}
\end{eqnarray}%
where $s_{i}$ and $t_{i}$ are constants. They can be determined by
evaluating the divergent parts of the diagrams shown in fig. \ref{grafici},
where the wiggled line denotes the field $h_{\mu \nu }$, the continuous line
with the arrow denotes the Faddeev-Popov ghosts and the double lines denote
either the sources $K_{g}$ coupled to the $g_{\mu \nu }$ transformations or
the sources $K_{C}$ coupled to the $C$ transformations. 

The calculation proceeds as follows.

1) Using (\ref{fieldredef}), the terms proportional to $K_{\rho }^{C}$
contained in $S_{\text{count}}$ at $h_{\mu \nu }=0$ are the integral of%
\begin{equation*}
-\Delta \mathcal{R}^{\rho }K_{\rho }^{C}=\kappa ^{2}\left[ \left( s_{1}+%
\frac{s_{2}}{2}\right) C^{\tau }(\partial _{\tau }C^{\rho })+\frac{s_{2}}{2}%
C^{\tau }(\partial ^{\rho }C_{\tau })+s_{3}C^{\rho }(\partial _{\tau
}C^{\tau })\right] K_{\rho }^{C}.
\end{equation*}%
We can work out the values of the constants $s_{i}$ by computing the first
diagram of fig. \ref{grafici}, which gives%
\begin{equation}
\Delta C^{\rho }=\frac{\kappa ^{2}}{12(4\pi )^{2}\alpha \lambda \omega
^{2}\varepsilon }\left[ 3C^{\rho }-8\kappa (2\omega ^{2}+1)h_{\mu }^{\rho
}C^{\mu }+\kappa (4\omega ^{2}-1)hC^{\rho }\right] .  \label{deltaC}
\end{equation}

2) The terms proportional to $K^{\mu \nu }$ contained in $S_{\text{count}}$
are equal to the integral of%
\begin{eqnarray}
&&-\Delta \mathcal{R}_{\mu \nu }K^{\mu \nu }=\kappa
^{2}[(t_{1}-2s_{1})K^{\mu \nu }\partial _{\mu }C_{\nu }+t_{2}K\partial \cdot
C]+2\kappa ^{3}(\partial _{\mu }K^{\mu \nu })(s_{2}h_{\nu \rho }C^{\rho
}+s_{3}hC_{\nu })  \notag \\
&&\qquad +\kappa ^{3}[(t_{3}-4s_{1})K^{\mu \nu }h_{\rho \mu }\partial _{\nu
}C^{\rho }-2s_{1}K^{\mu \nu }C^{\rho }\partial _{\rho }h_{\mu \nu
}+t_{3}K_{\nu }^{\mu }h_{\mu }^{\rho }\partial _{\rho }C^{\nu }+t_{4}K^{\mu
\nu }h_{\mu \nu }\partial \cdot C  \notag \\
\qquad &&\qquad +(t_{4}-2t_{2})K^{\mu \nu }h\partial _{\mu }C_{\nu
}+2(t_{5}+t_{2})Kh_{\mu }^{\rho }\partial _{\rho }C^{\mu }+2t_{6}Kh\partial
\cdot C]  \label{deltark}
\end{eqnarray}%
to the first order in $h_{\alpha \beta }$, where $K\equiv K_{\mu }^{\mu }$.
We can work out the terms of order $\kappa ^{2}$ by computing the second
diagram of fig. \ref{grafici}, which gives%
\begin{eqnarray}
t_{1}\pi ^{2}\varepsilon &=&-\frac{5}{18\alpha }-\frac{1}{3\alpha \lambda }-%
\frac{1}{24\alpha \lambda \omega ^{2}}+\frac{5}{18\alpha \omega }+\frac{1}{%
9\xi }+\frac{1}{12\xi \omega ^{2}}+\frac{5}{36\xi \omega },  \notag \\
t_{2}\pi ^{2}\varepsilon &=&\frac{5}{72\alpha }-\frac{5}{48\alpha \lambda }-%
\frac{1}{192\alpha \lambda \omega ^{2}}-\frac{5}{72\alpha \omega }-\frac{1}{%
36\xi }-\frac{1}{48\xi \omega ^{2}}-\frac{5}{144\xi \omega }.  \label{t12}
\end{eqnarray}

3) The other coefficients $t_{i}$ are obtained by computing the third and
fourth diagrams of fig. \ref{grafici}, which give%
\begin{eqnarray}
t_{3}\pi ^{2}\varepsilon &=&-\frac{25}{72\alpha }-\frac{1}{12\alpha \lambda }%
-\frac{5}{72\alpha \omega ^{2}}+\frac{1}{48\alpha \lambda \omega ^{2}}+\frac{%
5}{12\alpha \omega }+\frac{7}{72\xi }+\frac{7}{144\xi \omega ^{2}}+\frac{1}{%
12\xi \omega },  \notag \\
t_{4}\pi ^{2}\varepsilon &=&\frac{25}{144\alpha }+\frac{5}{144\alpha \omega
^{2}}-\frac{5}{24\alpha \omega }-\frac{7}{144\xi }-\frac{7}{288\xi \omega
^{2}}-\frac{1}{24\xi \omega },  \notag \\
t_{5}\pi ^{2}\varepsilon &=&\frac{5}{32\alpha }+\frac{5}{24\alpha \lambda }+%
\frac{25}{288\alpha \omega ^{2}}+\frac{1}{96\alpha \lambda \omega ^{2}}+%
\frac{25}{144\alpha \omega }-\frac{1}{96\xi }-\frac{11}{576\xi \omega ^{2}}-%
\frac{7}{144\xi \omega },  \notag \\
t_{6}\pi ^{2}\varepsilon &=&-\frac{35}{576\alpha }-\frac{5}{192\alpha \omega
^{2}}-\frac{5}{288\alpha \omega }+\frac{5}{576\xi }+\frac{1}{128\xi \omega
^{2}}+\frac{5}{288\xi \omega }.  \label{t3456}
\end{eqnarray}

4)\ A separate discussion concerns the coefficient $t_{0}$, which may be
seen as the renormalization of the flat-space background metric $\eta _{\mu
\nu }$ (after a redefinition of $t_{1}$). Observe that the contribution $%
\kappa ^{2}t_{0}g_{\mu \nu }$ to the field redefinition $\Delta g_{\mu \nu }$
drops out of (\ref{deltark}), because it is covariant. It only adds%
\begin{equation}
\kappa ^{2}t_{0}\int \frac{\delta S_{\text{HD}}}{\delta g_{\mu \nu }}g_{\mu
\nu }  \label{to}
\end{equation}%
to the first term on the right-hand side of equation (\ref{sf}). Since (\ref%
{to}) is proportional to the field equations, its coefficient $t_{0}$ may be
gauge dependent. However, (\ref{to}) is also covariant, so it may mix with
the renormalizations of $\Lambda _{C}$, $\zeta $, $\alpha $ and $\xi $. This
means that only the combinations of such coefficients that are not affected
by $t_{0}$ are truly gauge independent. It is simple to check that such
combinations are $\Lambda _{C}/\zeta ^{2}$, $\alpha $ and $\xi $. For
convenience, we define 
\begin{equation}
t_{0}\pi ^{2}\varepsilon =\frac{3}{16\alpha \lambda }+\frac{1}{64\alpha
\lambda \omega ^{2}}-\frac{3}{64\xi \omega ^{2}}-\frac{3}{16\xi \omega }+%
\frac{A}{8},  \label{t0}
\end{equation}%
where $A$ is an arbitrary constant that parametrizes the surviving gauge
dependence.

5) The coefficients $\Delta \Lambda _{C}$, $\Delta \zeta $, $\Delta \alpha $
and $\Delta \xi $ of $S_{\text{count}}$ can be worked out by computing the
graviton self energy. We obtain%
\begin{eqnarray}
\Delta \alpha &=&-\frac{133}{10},\qquad \Delta \xi =\frac{5}{6}+\frac{5\xi }{%
\alpha }+\frac{5\xi ^{2}}{3\alpha ^{2}},\qquad \Delta \zeta =\zeta \left( 
\frac{5}{6\xi }+\frac{5\xi }{3\alpha ^{2}}+A\right) ,  \notag \\
\Delta \Lambda _{C} &=&\Lambda _{C}\left( -\frac{5}{\alpha }+\frac{2}{\xi }%
-2A\right) -\frac{5\zeta ^{2}}{4\alpha ^{2}}-\frac{\zeta ^{2}}{4\xi ^{2}}.
\label{deltas}
\end{eqnarray}%
As promised, the combinations $\zeta \Delta \Lambda _{C}-2\Lambda _{C}\Delta
\zeta $, $\Delta \alpha $ and $\Delta \xi $ are independent of the arbitrary
constant $A$.

The beta functions are%
\begin{equation}
\beta _{\alpha }=-\frac{2\kappa ^{2}}{(4\pi )^{2}}\Delta \alpha ,\qquad
\beta _{\xi }=-\frac{2\kappa ^{2}}{(4\pi )^{2}}\Delta \xi ,\qquad \beta
_{\zeta }=-\frac{2\kappa ^{2}}{(4\pi )^{2}}\Delta \zeta ,\qquad \beta
_{\Lambda _{C}}=-\frac{2\kappa ^{2}}{(4\pi )^{2}}\Delta \Lambda _{C}.
\label{betas}
\end{equation}

Now we compare our results with those of the literature. To our knowledge,
the values of the coefficients $s_{i}$, $i=1,2,3$, encoded in formula (\ref%
{deltaC}), and those of the coefficients $t_{i}$, $i=3,4,5,6$, of formula (%
\ref{t3456}) were not known. The most complete results for the other
quantities are those of Salvio and Strumia, collected in ref. \cite%
{agravity2}. The notation of that paper is related to ours by the
redefinitions%
\begin{equation*}
\alpha =\frac{2}{f_{2}^{2}},\qquad \xi =\frac{2}{f_{0}^{2}},\qquad \zeta =%
\bar{M}_{\text{Pl}}^{2},\qquad \Lambda _{C}=\Lambda ,\qquad \omega =\frac{%
c_{g}}{2}-1,\qquad \lambda =-\frac{f_{2}^{2}}{\xi _{g}},
\end{equation*}%
together with $\sigma =0$, $\kappa =1$, while our $h_{\mu \nu }$ is equal to
the one of \cite{agravity2} divided by two. Formulas (\ref{betas}) and the
coefficients $t_{1}$ and $t_{2}$ of equations (\ref{t12})\ agree with those
of \cite{agravity2}, apart from the following discrepancies: the term $%
-5f_{2}^{4}/(3f_{0}^{4})\bar{M}_{\text{Pl}}^{2}$ in formula (54a) of \cite%
{agravity2} should be replaced by $-5f_{2}^{4}/(3f_{0}^{2})\bar{M}_{\text{Pl}%
}^{2}$; moreover, the right-hand side of formula (55) should be multiplied
by an overall minus sign and its first fraction should be multiplied by an
extra factor $1/(c_{g}-2)$. The authors of \cite{agravity2} have implicitly
set $t_{0}=0$. This choice determines the constant $A$, which is related to
the constant $X$ of \cite{agravity2} by the equation $A=-X-3f_{0}^{2}/4$.

Finally, we make a nontrivial check of $\Delta g_{\mu \nu }$ by adding $%
N_{s} $ scalar fields with the minimal action%
\begin{equation}
S_{\text{s}}=\frac{1}{2}\sum_{i=1}^{N_{s}}\int \sqrt{-g}g^{\mu \nu
}(\partial _{\mu }\varphi ^{i})(\partial _{\nu }\varphi ^{i}).  \label{sscal}
\end{equation}%
The total action $S_{\text{HD}}+S_{\text{s}}$ is renormalizable. Indeed, the
external scalar legs of every diagram carry derivatives, so the vertex $%
\sqrt{-g}\varphi ^{4}$ and the nonminimal term $\sqrt{-g}R\varphi ^{2}$ are
not generated as counterterms, if they are absent at the classical level. No
other counterterms are compatible with power counting and invariance under
diffeomorphisms.

We find that the scalar self energy and the scalar-graviton vertex are
renormalized by the field redefinitions $\Delta g_{\mu \nu }$ found above
plus 
\begin{equation}
\Delta \varphi ^{i}=-\frac{\kappa ^{2}}{(4\pi )^{2}\varepsilon }\left( A+%
\frac{3}{2\xi }\right) \varphi ^{i}.  \label{deltaphi}
\end{equation}

\section{Absorptive part}

\setcounter{equation}{0} \label{abso}

In this section we calculate the absorptive part of the graviton self
energy. We work in the limit $\Lambda _{C}=0$ and include $N_{s}$ scalar
fields $\varphi ^{i}$ coupled to gravity by the action (\ref{sscal}). At the
end, we add other types of matter fields. The calculation gives us the
chance to show that the graviton/fakeon prescription is consistent and leads
to physical predictions that are different from those obtained by quantizing
the classical action (\ref{lhd}) by means of standard prescriptions \cite%
{backfm,percacci,agravity,agravity2,stelle}.

For simplicity, we set the gauge-fixing parameters $\lambda $ and $\sigma $
to one, but keep $\omega $ arbitrary to check that the physical quantities
we compute are gauge independent. However, due to the complications of some
formulas, we report the gauge-dependent results only for $\omega =-1/2$
(which is the de Donder gauge). The results for arbitrary $\omega $ can be
downloaded from \href{http://renormalization.com/Math/QG}{the link} \cite%
{betaQG}, together with the Mathematica programs used for the calculations
of this paper.

It is convenient to integrate $B_{\mu }$ out in formula (\ref{skpsi}), which
is equivalent to replacing it with the solution%
\begin{equation*}
B_{\mu }=\frac{1}{2\kappa ^{2}}\mathcal{G}_{\mu }
\end{equation*}%
of its own field equation, or making the replacement%
\begin{equation*}
(S_{K},\Psi )\rightarrow \frac{\mu ^{-\varepsilon }}{4\kappa ^{2}}\int 
\mathcal{G}^{\mu }\left( \zeta +\alpha \square \right) \mathcal{G}_{\mu }+S_{%
\text{gh}}
\end{equation*}%
in formula (\ref{lgr}). The free propagator of the metric fluctuation $%
h_{\mu \nu }$ reads%
\begin{equation}
\langle h_{\mu \nu }(p)\hspace{0.01in}\hspace{0.01in}h_{\rho \sigma
}(-p)\rangle _{0}=\frac{i\mathcal{I}_{\mu \nu \rho \sigma }}{2p^{2}(\zeta
-\alpha p^{2})}+\frac{i(\alpha -\xi )\varpi _{\mu \nu }\varpi _{\rho \sigma }%
}{6(p^{2})^{2}(\zeta -\alpha p^{2})(\zeta -\xi p^{2})}  \label{propo}
\end{equation}%
at $\omega =-1/2$, where%
\begin{equation*}
\mathcal{I}_{\mu \nu \rho \sigma }=\eta _{\mu \rho }\eta _{\nu \sigma }+\eta
_{\mu \sigma }\eta _{\nu \rho }-\eta _{\mu \nu }\eta _{\rho \sigma },\qquad
\varpi _{\mu \nu }=p^{2}\eta _{\mu \nu }+2p_{\mu }p_{\nu }.
\end{equation*}

We define the \textit{graviton/fakeon prescription} by introducing two
widths $\epsilon $ and $\mathcal{E}$ as follows:

($a$) replace $p^{2}$ with $p^{2}+i\epsilon $ everywhere in the denominators
of the propagators;

($b$) turn the massive poles into fakeons by means of the replacement%
\begin{equation*}
\frac{1}{\zeta -u(p^{2}+i\epsilon )}\rightarrow \frac{\zeta -up^{2}}{(\zeta
-u(p^{2}+i\epsilon ))^{2}+\mathcal{E}^{4}},
\end{equation*}%
where $u$ is equal to $\alpha $ or $\xi $.

$c$) calculate the diagrams in the Euclidean framework, nonanalytically Wick
rotate them as explained in refs. \cite{fakeons,LWformulation,LWunitarity},
then make $\epsilon $ tend to zero first and $\mathcal{E}$ tend to zero last.

It is convenient to apply the prescription after separating the graviton
poles from the fakeon poles by means of a partial fractioning. Specifically,
we use formulas such as%
\begin{equation}
\frac{1}{z(1-az)}=\frac{1}{z}+\frac{a}{1-az},\qquad \frac{1}{%
z^{2}(1-az)(1-bz)}=\frac{1}{z^{2}}+\frac{a+b}{z}+\frac{1}{a-b}\left( \frac{%
a^{3}}{1-az}-\frac{b^{3}}{1-bz}\right) ,  \label{affake}
\end{equation}%
etc., with $z=p^{2}$, $a=\alpha /\zeta $ and $b=\xi /\zeta $, to decompose
the propagator (\ref{propo}) as the sum 
\begin{equation}
\langle h_{\mu \nu }(p)\hspace{0.01in}\hspace{0.01in}h_{\rho \sigma
}(-p)\rangle _{0}=\langle h_{\mu \nu }(p)\hspace{0.01in}\hspace{0.01in}%
h_{\rho \sigma }(-p)\rangle _{0\text{\hspace{0.01in}grav}}+\langle h_{\mu
\nu }(p)\hspace{0.01in}\hspace{0.01in}h_{\rho \sigma }(-p)\rangle _{0\text{%
\hspace{0.01in}fake}}  \label{decompro}
\end{equation}%
of a graviton part plus a fake part, where the graviton part collects the
poles at $p^{2}=0$, while the fake part collects the poles at $p^{2}=\zeta
/\alpha $ and $p^{2}=\zeta /\xi $. Once we apply the graviton/fakeon
prescription as explained above, we obtain%
\begin{eqnarray*}
\langle h_{\mu \nu }(p)\hspace{0.01in}\hspace{0.01in}h_{\rho \sigma
}(-p)\rangle _{0\text{\hspace{0.01in}grav}} &=&\frac{i}{2\zeta
(p^{2}+i\epsilon )}\left[ \mathcal{I}_{\mu \nu \rho \sigma }+\frac{(\alpha
-\xi )\varpi _{\mu \nu }\varpi _{\rho \sigma }}{3\zeta ^{2}}\left( \frac{%
\zeta }{p^{2}+i\epsilon }+\alpha +\xi \right) \right] , \\
\langle h_{\mu \nu }(p)\hspace{0.01in}\hspace{0.01in}h_{\rho \sigma
}(-p)\rangle _{0\text{\hspace{0.01in}fake}} &=&\frac{i\alpha \mathcal{I}%
_{\mu \nu \rho \sigma }(\zeta -\alpha p^{2})}{2\zeta \lbrack (\zeta -\alpha
(p^{2}+i\epsilon ))^{2}+\mathcal{E}^{4}]} \\
&&+\frac{i\varpi _{\mu \nu }\varpi _{\rho \sigma }}{6\zeta ^{3}}\left( \frac{%
\alpha ^{3}(\zeta -\alpha p^{2})}{(\zeta -\alpha (p^{2}+i\epsilon ))^{2}+%
\mathcal{E}^{4}}-\frac{\xi ^{3}(\zeta -\xi p^{2})}{(\zeta -\xi
(p^{2}+i\epsilon ))^{2}+\mathcal{E}^{4}}\right) .
\end{eqnarray*}

We compute the absorptive part of the graviton self energy at $\Lambda
_{C}=0 $. Recall that the absorptive part of an amplitude is equal to its
imaginary part, so the one of a diagram is equal to minus its real part. The
calculation involves three bubble diagrams with external legs $h_{\mu \nu }$%
. The loop can be made of scalar fields, Faddeev-Popov ghosts or $h_{\mu \nu
}$ itself. The scalar contributions are not interested by the fakeons, so
they coincide with those of Einstein gravity. The same conclusion applies to
the contributions of the Faddeev-Popov ghosts, as is evident by working with
the action (\ref{ghaccon}). So, we focus on the bubble diagram of the metric
fluctuation $h_{\mu \nu }$.

Now we prove that the fakeons do not affect the real part of this diagram,
so we can drop them and replace the propagators (\ref{propo}) with $\langle
h_{\mu \nu }\hspace{0.01in}\hspace{0.01in}h_{\rho \sigma }\rangle _{0\text{%
\hspace{0.01in}grav}}$. The $h_{\mu \nu }$ bubble diagram obviously contains
two propagators. Decomposing each of them as shown in formula (\ref{decompro}%
), we obtain the sum of three terms: ($i$) the pure graviton contributions,
where each propagator is replaced by its graviton part $\langle h_{\mu \nu }%
\hspace{0.01in}\hspace{0.01in}h_{\rho \sigma }\rangle _{0\text{\hspace{0.01in%
}grav}}$; ($ii$) the pure fakeon contributions, where each propagator is
replaced by its fakeon part $\langle h_{\mu \nu }\hspace{0.01in}\hspace{%
0.01in}h_{\rho \sigma }\rangle _{0\text{\hspace{0.01in}fake}}$; ($iii$) the
mixed contributions, where one propagator is replaced by its graviton part
and the other propagator is replaced by the fakeon part.

The contributions of type ($ii$) and ($iii$) can be dropped, because they
are purely imaginary. To see this, recall that the diagram must be
calculated in the Euclidean framework and then nonanalytically Wick rotated
as explained in refs. \cite{fakeons,LWformulation,LWunitarity}. Moreover, we
must first work at finite $\mathcal{E}$, let $\epsilon $ tend to zero while $%
\mathcal{E}$ is finite and nonzero, and finally let $\mathcal{E}$ also tend
to zero. Varying the energy $p^{0}$ of the external momentum $p$, the poles
of the propagators may pinch the integration domain. When $p^{0}$ is located
below the thresholds of the pinchings, the result of the loop integral is
purely imaginary. Indeed, an overall factor $i$ is brought by the residue
theorem, applied to the integral on the loop energy $k^{0}$. After that, the 
$i\epsilon $ prescription is redundant below the thresholds, which allows us
to let $\epsilon \rightarrow 0$ at the level of the integrand. Since the
integrand is real in this limit, the result of the integral is purely
imaginary.

In case ($iii$), no threshold is located on the real $p^{0}$ axis for $%
\mathcal{E}>0$, $\epsilon \rightarrow 0$, since the pinchings occur far away
(at a distance roughly equal to $\mathcal{E}$). Then, the Wick rotation is
analytic for real $p$ and the result is purely imaginary (for every $%
\mathcal{E}>0$ and so also when $\mathcal{E}$ tends to zero).

In case ($ii$) the $i\epsilon $ prescription is redundant from the
beginning, because only the fakeons circulate in the loop. Some thresholds
of the pinchings lie on the real axis of the complex $p^{0}$ plane. Again,
the result is purely imaginary below the thresholds. We can reach the
regions above the thresholds by means of the \textit{average continuation} 
\cite{fakeons,LWformulation,LWunitarity}, which is the arithmetic average of
the two analytic continuations that circumvent the thresholds. Clearly, the
average continuation of a function that is purely imaginary in a real
interval of the complex $p^{0}$ plane, is purely imaginary on the entire
real $p^{0}$ axis.

In conclusion, we can concentrate on the contributions of type ($i$), which
can be evaluated by using $\langle h_{\mu \nu }\hspace{0.01in}\hspace{0.01in}%
h_{\rho \sigma }\rangle _{0\text{\hspace{0.01in}grav}}$ as the propagator of 
$h_{\mu \nu }$. If we make some further steps, we can prove that the
surviving contributions are uniquely determined by the divergent part of the
graviton self energy, calculated in the low-energy expansion, which means
expanding the action $S_{\text{HD}}$ of formula (\ref{lhd}) around the
Hilbert term $\int \sqrt{-g}R$ and treating the parameters $\alpha $ and $%
\xi $ perturbatively. At the same time, the result obtained with this method
is exact in $\alpha $ and $\xi $.

To show these properties, we first make some observations about the
expansion in question. It can be worked out by starting from the self energy
diagram studied in the previous section and expanding its integrand in
powers of $\alpha $ and $\xi $. Since the vertices depend on such parameters
polynomially, it is sufficient to concentrate on the expansions of the
propagators. When we expand the propagator (\ref{propo}), or its $\omega
\neq -1/2$ version, we just obtain poles at $p^{2}=0$. It is sufficient to
truncate the expansion of the propagator to the quadratic order in $\alpha $
and $\xi $, because higher powers simplify the poles and just multiply
polynomials of the momentum. Inside the bubble diagram, these corrections
give massless tadpoles (polynomials times a single massless propagator) and
are set to zero by the dimensional regularization.

It is obvious [and easy to check, using formulas of type (\ref{affake})]
that the propagator, once truncated to the quadratic order in $\alpha $ and $%
\xi $, coincides with $\langle h_{\mu \nu }\hspace{0.01in}\hspace{0.01in}%
h_{\rho \sigma }\rangle _{0\text{\hspace{0.01in}grav}}$, up to polynomials,
which, again, are negligible for our purposes. Thus, the absorptive part of
the graviton self energy can be calculated by means of the low-energy
expansion. It remains to show that it is uniquely determined by the
divergent part.

Since no parameters of positive dimensions in units of mass are present (the
cosmological constant being set to zero), the result of the loop integral,
calculated by expanding the integrand in $\alpha $ and $\xi $, must be a
polynomial times $\ln (-p^{2})$, where $p$ is the external momentum. Then it
is clear that the divergent part and the absorptive part of the diagram are
unambiguously related to each other. A quick way to see this is by means of
the chain of relations 
\begin{equation}
\frac{1}{\varepsilon }\rightarrow \frac{1}{2}\ln \Lambda ^{2}\rightarrow 
\frac{1}{2}\ln \frac{\Lambda ^{2}}{-p^{2}}\rightarrow -\frac{1}{2}\ln
(-p^{2})\overset{\text{prescr}}{\longrightarrow }-\frac{1}{2}\ln
(-p^{2}-i\epsilon )\overset{\text{abs}}{\longrightarrow }i\frac{\pi }{2}%
\theta (p^{2}).  \label{repla}
\end{equation}%
The first arrow relates the poles of the dimensional regularization to the
logarithms of an ordinary cutoff $\Lambda $. The second and third arrow
relate them to the logarithms of the external momentum $p$. The fourth arrow
restores the Feynman prescription (which is the only prescription to be used
at this point, since no fakeons have survived). The last arrow extracts the
contribution to the absorptive part.

To summarize, the absorptive part of the graviton self energy can be
calculated from the divergent part of the expansion in powers of $\alpha $
and $\xi $ and the result is exact in $\alpha $ and $\xi $. Thanks to this,
the outcome is guaranteed to be gauge invariant (after applying field
redefinitions and procedures analogous to the ones described in the previous
section, adapted to the power counting of the low-energy expansion).

Note that we have slightly modified the prescription given ref. \cite{LWgrav}
to make gauge invariance manifest. Strictly speaking, the widths $\epsilon $
and $\mathcal{E}$ break gauge invariance, which must be recovered in the
limit $\epsilon \rightarrow 0$ followed by $\mathcal{E}\rightarrow 0$. In
general, it might be necessary to add corrections proportional to $\epsilon $
and/or $\mathcal{E}$ to implement the recovery of gauge invariance. The
graviton/fakeon prescription formulated in this section is optimized to make
this extra effort unnecessary.

Let us comment on how the chain of relations (\ref{repla}) is modified in
the case of the fakeons. There we have%
\begin{equation}
\frac{1}{\varepsilon }\rightarrow \frac{1}{2}\ln \Lambda ^{2}\rightarrow 
\frac{1}{2}\ln \frac{\Lambda ^{2}}{-p^{2}}\rightarrow -\frac{1}{2}\ln
(-p^{2})\overset{\text{prescr}}{\longrightarrow }-\frac{1}{4}\ln (-p^{2})^{2}%
\overset{\text{abs}}{\longrightarrow }0,  \label{chainfake}
\end{equation}%
so no absorptive part survives. The explanation of the fourth arrow can be
found in ref. \cite{LWgrav} and amounts to the fakeon prescription. In
practice, the ultraviolet behavior of a two-point function is governed by
two types of logarithms of the momentum. One, $\ln (-p^{2}-i\epsilon )$, is
inherited from the Feynman prescription, which is associated with the
physical degrees of freedom. The other one, $(1/2)\ln (p^{2})^{2}$, is
inherited from the fakeon prescription. From the point of view of the
ultraviolet divergences, they are both equal to $-2/\varepsilon $, which is
why we had to use the tricks described above to disentangle them. Their
difference gives the absorptive part, due to the identity%
\begin{equation}
-\ln (-p^{2}-i\epsilon )+\frac{1}{2}\ln (p^{2})^{2}=i\pi \theta (p^{2}).
\label{absorlog}
\end{equation}

Because of the fakeons, the imaginary and real parts of a loop diagram are
unrelated to each other. The divergent part obeys the renormalizable power
counting of the higher-derivative theory, while the absorptive part obeys
the (nonrenormalizable) power counting of the low-energy expansion and is
consistent with unitarity. Contributions of higher dimensions (multiplied by
large powers of $\alpha $ and $\xi $) can appear in the absorptive part,
multiplied by either side of (\ref{absorlog}), without affecting the
divergent part. This is the basic mechanism by means of which the fakeons
make renormalization and unitarity compatible with each other, in the limit
of vanishing cosmological constant.

At this point, the calculation is straightforward. With the help of field
redefinitions of the form%
\begin{eqnarray}
\Delta g_{\mu \nu } &=&\kappa ^{3}\frac{i\pi }{2}\theta (-\square )\left[
-2a_{1}\square h_{\mu \nu }-a_{2}\eta _{\mu \nu }\square h-a_{3}\eta _{\mu
\nu }\partial ^{\rho }\partial ^{\sigma }h_{\rho \sigma }-a_{4}\partial
_{\mu }\partial _{\nu }h\right.  \notag \\
&&-\left. 2a_{5}(\partial _{\mu }\partial ^{\rho }h_{\rho \nu }+\partial
_{\nu }\partial ^{\rho }h_{\rho \mu })+a_{6}\partial _{\mu }\partial _{\nu
}\partial ^{\rho }\partial ^{\sigma }h_{\rho \sigma }\right] ,
\label{deltacca}
\end{eqnarray}%
where $a_{i}$, $i=1,\ldots 6$, are functions of $\square $, the absorptive
part of the graviton self energy is encoded into the contribution 
\begin{equation}
\Gamma _{\text{abs}}=\frac{iN_{s}\mu ^{-\varepsilon }}{120(16\pi )}\int 
\sqrt{-g}\left[ R_{\mu \nu }\theta (-\square _{c})R^{\mu \nu }+\frac{1}{2}%
R\theta (-\square _{c})R\right] -\int \frac{\delta S_{\text{HD}}}{\delta
g_{\mu \nu }}\Delta g_{\mu \nu }  \label{gab}
\end{equation}%
to the functional $\Gamma $, where $\square _{c}=g^{\rho \sigma }D_{\rho
}D_{\sigma }$ is the covariant D'Alembertian, $D_{\rho }$ being the
covariant derivative. The coefficients of the field redefinitions (\ref%
{deltacca}) are rather lengthy, so we just report them in the simple case $%
\omega =-1/2$: 
\begin{eqnarray}
2160a_{1}(4\pi )^{2}\zeta ^{6} &=&-\alpha (\alpha ^{2}-\xi ^{2})^{2}\square
^{5}-\zeta (\alpha -\xi )^{2}(23\alpha ^{2}+24\alpha \xi +\xi ^{2})\square
^{4}  \notag \\
&&-2\zeta ^{2}(115\alpha ^{3}-99\alpha ^{2}\xi -27\alpha \xi ^{2}+11\xi
^{3})\square ^{3}  \notag \\
&&+4\zeta ^{3}(89\alpha ^{2}-106\alpha \xi +17\xi ^{2})\square ^{2}+72\zeta
^{4}(2\alpha +\xi )\square -4392\zeta ^{5},  \notag \\
3240a_{2}(4\pi )^{2}\zeta ^{6} &=&(\alpha +5\xi )(\alpha ^{2}-\xi
^{2})^{2}\square ^{5}+4\zeta (\alpha -\xi )^{2}(7\alpha ^{2}+6\alpha \xi
-\xi ^{2})\square ^{4}  \notag \\
&&+4\zeta ^{2}(\alpha -\xi )^{2}(55\alpha +38\xi )\square ^{3}-48\zeta
^{3}(7\alpha ^{2}-53\alpha \xi +46\xi ^{2})\square ^{2}  \notag \\
&&+432\zeta ^{4}(3\alpha +29\xi )\square +9072\zeta ^{5},  \notag \\
9(a_{2}+a_{3})(4\pi )^{2}\zeta ^{6} &=&-2\zeta ^{4}(\alpha -\xi )\square .
\label{deltag}
\end{eqnarray}%
Their $\omega $-dependent expressions can be found at \href{http://renormalization.com/Math/QG}%
{the link} \cite{betaQG}. The other coefficients of the field redefinition
remain undetermined.

Adding (massless) matter fields of all types and using the results of refs. 
\cite{hathrell}, formula (\ref{gab}) turns into%
\begin{eqnarray}
\Gamma _{\text{abs}} &=&\frac{i\mu ^{-\varepsilon }}{16\pi }\int \sqrt{-g}%
\left[ c\left( R_{\mu \nu }\theta (-\square _{c})R^{\mu \nu }-\frac{1}{3}%
R\theta (-\square _{c})R\right) +\frac{N_{s}\eta ^{2}}{36}R\theta (-\square
_{c})R\right]  \notag \\
&&-\int \frac{\delta S_{\text{HD}}}{\delta g_{\mu \nu }}\Delta g_{\mu \nu },
\label{gabs}
\end{eqnarray}%
where $\Delta g_{\mu \nu }$ is unmodified,%
\begin{equation*}
c=\frac{1}{120}(N_{s}+6N_{f}+12N_{v})
\end{equation*}%
is known as \textquotedblleft central charge\textquotedblright\ ($N_{f}$
being the numbers of Dirac fermions plus one half the number of Weyl
fermions and $N_{v}$ \ being the number of massless vectors) and $\eta $ is
related to the coefficient of the nonminimal coupling of the scalar fields,
obtained by extending (\ref{sscal}) into%
\begin{equation*}
S_{s}=\frac{1}{2}\sum_{i=1}^{N_{s}}\int \sqrt{-g}\left[ g^{\mu \nu
}(\partial _{\mu }\varphi ^{i})(\partial _{\nu }\varphi ^{i})+\frac{1}{6}%
(1+2\eta )R\varphi ^{i\hspace{0.01in}2}\right] .
\end{equation*}

Note that formula (\ref{gabs}) is nonlocal, because it is the convergent
part of an amplitude. Since the nonlocality is just due to the $\theta $
function, we can call it \textit{piecewise local}. We recall that the
amplitudes satisfy nonlocal WTST\ identities, encoded into the $\Gamma $
master equation (\ref{mastg}). Although the field redefinitions and the
symmetry transformations involved in such identities are nonlocal, their
nonlocalities are under control, because they are generated by other kinds
of amplitudes. See \cite{ABWard} for details on the general theory and
references.

We stress again that formulas (\ref{gab}), (\ref{deltag}) and (\ref{gabs})
are exact in $\alpha $ and $\xi $, even if we worked them out by means of an
expansion. The results (\ref{gab}) and (\ref{gabs}) are gauge invariant and
gauge independent, as they should, apart from the last term, which vanishes
on the solutions of the $S_{\text{HD}}$ field equations. In particular, we
have verified that every dependence on $\omega $ can be absorbed into a
suitable $\Delta g_{\mu \nu }$. On the other hand, the contributions of the
matter fields cannot be absorbed into a piecewise local redefinition $\bar{%
\Delta}g_{\mu \nu }$ of the metric tensor, because they do not vanish on the
solutions of the $S_{\text{HD}}$ field equations.

It is worth to point out that when the Feynman prescription is used for all
the poles of the free propagators, which is what is done in the ordinary
approaches \cite{backfm,percacci,agravity,agravity2,stelle}, the absorptive
part of the graviton self energy receives nontrivial contributions from the
spin-2 ghosts. This proves that the graviton/fakeon prescription leads to a
different theory.

Finally, we can check that the physical degrees of freedom are indeed the
graviton and the matter fields by showing that the graviton self energy
satisfies the correct optical theorem. At the perturbative level, the
optical theorem and the unitarity equation $SS^{\dag }=1$ are encoded into
the so-called cutting equations \cite{cuttingeq}. Since the absorptive part
of the graviton self energy is determined by the low-energy expansion, it
satisfies the cutting equations of that expansion, which are consistent with
unitarity at vanishing cosmological constant \cite{unitarity}. Then, the cut
propagators of the complete theory, which encode the physical spectrum,
coincide with those of the low-energy expansion, which are determined by the
Hilbert term and the matter action. Thus, they receive contributions from
the graviton and the matter fields, but not the fakeons.

\section{Conclusions and outlook}

\label{concl} \setcounter{equation}{0}

In this paper we have studied the theory of quantum gravity proposed in ref. 
\cite{LWgrav}, by computing its renormalization at one loop and the
absorptive part of the graviton self energy. The theory is the unique
strictly renormalizable one of a larger class of theories, where the ghosts
are eliminated by turning the poles of the free propagators that are due to
the higher derivatives into fakeons. The fakeons are degrees of freedom that
contribute to the correlation functions (to the extent that they make the
theory renormalizable) but disappear from the physical spectrum, saving
perturbative unitarity.

The renormalization coincides with the one of the Euclidean version of the
theory and the results we have found are consistent with those that can be
found the literature. Without making use of the background field method, we
managed to save the calculation of the diagrams with three external graviton
legs by computing the renormalization of the symmetry transformations and
using the WTST\ identities. We have extended the results available in the
literature by computing the first nonlinear corrections to the field
renormalizations of the metric tensor and the Faddeev-Popov ghosts.

The absorptive part of the graviton self energy is a key quantity to
appreciate the crucial differences between the theory of quantum gravity
studied here and other quantizations of the same classical action. At zero
cosmological constant, a number of tricks allow us to relate it to the
renormalization of the theory expanded around the Hilbert term. The final
result is the sum of a term proportional to the central charge $c$ of the
matter fields coupled to gravity, plus a term due to the nonminimal coupling
of the scalar fields, plus corrections that vanish on the solutions of the
field equations. The correct optical theorem is satisfied, with no
contributions from the fakeons.

We conclude by mentioning some interesting outlooks. With some additional
effort, the calculation of the absorptive part can be extended to $\Lambda
_{C}\neq 0$. However, contributions similar to those of massive tadpoles
(which are divergent, but have no absorptive part) are present, so the
relation (\ref{repla}) cannot be applied straightforwardly. Since the free
propagator $\langle h_{\mu \nu }h_{\rho \sigma }\rangle _{0}$ has a massive
scalar pole with a positive residue, another interesting possibility is to
let the theory propagate an additional massive scalar field, which was
implicitly turned into a fakeon in this paper.

\end{document}